\documentclass[twocolumn,preprintnumbers,amsmah,amssymb,showpacs]{revtex4}
\usepackage{graphicx}
\usepackage{dcolumn}
\usepackage{bm}
 \usepackage[latin1]{inputenc}
\usepackage{multirow}
\usepackage{amsmath}
\usepackage{appendix}

\begin{document}                

\title{Polarized neutron channeling as a tool for the investigations of weakly magnetic thin films}

\author{S.V. Kozhevnikov$^{1,2}$,Yu.N. Khaydukov$^{3,4}$, T. Keller$^{3,4}$, F. Ott$^{2}$,  and F. Radu$^5$}
\affiliation{$^{1}$Frank Laboratory of Neutron Physics, Joint Institute for Nuclear Research, 141980 Dubna, Russian Federation }
\affiliation{$^{2}$Laboratoire L\'eon Brillouin UMR12 CEA/CNRS, F-91191 Gif sur Yvette, France}
\affiliation{$^{3}$Max-Planck-Institut f\"ur Festk\"orperforschung, Heisenbergstr. 1, D-70569 Stuttgart, Germany}
\affiliation{$^{4}$Max Planck Society Outstation at FRM-II, D-85747 Garching, Germany}
\affiliation{$^{5}$Helmholtz-Zentrum Berlin f\"ur Materialien und Energie, Albert-Einstein Strasse 15, D-12489, Berlin, Germany}
\date{\today}

\begin{abstract}We present and apply a new method to measure directly weak magnetization in thin films. The polarization of a neutron beam channeling through a thin film structure is measured after exiting the structure edge as a microbeam. We have applied the method to a tri-layer thin film structure acting as a planar waveguide for polarized neutrons. The middle guiding layer is a rare earth based ferrimagnetic material TbCo$_5$ with a low magnetization of about 20 mT. We demonstrate that the channeling method is more sensitive than the specular neutron reflection method.
\end{abstract}
\maketitle

Magnetic nanomaterials are widely used in practical applications and in fundamental physics. Therefore the development of new methods providing enhanced sensitivity for the characterization of their magnetic properties is a challenge. One of the well-established methods for magnetic thin films and nanostructures characterization is the Polarised Neutron Reflectometry (PNR) technique [1]. Its strength is the unique ability to probe the magnitude and the direction of the magnetization vector as a function of the depth in magnetic heterostructures. For collinear spin configurations, the PNR method resolves the depth dependence of the scattering length densities (SLD) $\rho(z)$ which depends on the neutron polarization: $\rho^{\pm}(z) = \rho_0(z) \pm \bf{c}B(z)$. Here $\rho_0(z)$ and $B(z)$ are the depth profiles of the nuclear SLD and the magnetic induction. By measuring reflectivities of spin-up and spin-down neutrons and fitting them to theoretical values one can extract the SLD depth profile and the depth dependence of the induction in a film. The method is routinely used for the characterization of ferromagnetic systems with rather high magnetization (about 0.3 - 2 T). Being a model dependent method it leads to the non-uniqueness of the resulting set of parameters describing the system. As a result, the accuracy of the fitted parameters depends both on the adequacy of a model and on the quality of the experimental data. For systems with weak magnetization ($\sim 10$~mT), PNR requires very long measurement times ($\sim 10^2$~ hours) in order to achieve sufficient statistics. This is usually not compatible with standard neutron reflectometry experiments. 

Weak magnetizations, typically less than 100 mT, are characteristic of ferrimagnetic materials or magnetic oxides. Magnetic materials such as Fe$_3$O4, CoFe$_2$O$_4$, BiFeO$_3$, GaFeO$_3$, Bi$_3$Fe$_5$O$_{12}$, NiCoMnAl are being increasingly studied because of their specific properties, either their electronic properties [2], their magneto-electric properties [3], their magneto-optical [4] or their magneto caloric properties [5]. The study of these materials is often limited by the weakness of the magnetic interfacial effects that are of interest. While XMCD [6] is a very sensitive technique to probe surface magnetism, it is sometimes rather difficult to apply to complex heterostructures as used in spintronic devices. Also, the XRMS [7] technique cannot access thicker layer structure because the soft x-ray penetration depth is on the order of 30 nm. The modelling of the XMCD and XRMS can also be rather complex in these non-metallic materials.

In this letter, we demonstrate that neutron channeling is a potentially new method to probe weak ferri- or ferromagnetic thin film structures.

The neutron channelling method both enhances the sensitivity to weak magnetization and reduces the model dependency. In order to perform a neutron channelling measurement, the weak magnetic layer B has to be sandwiched between two layers A and C with a higher SLD in order to form a resonator structure [8,9] (see Fig. 1a). Below the critical edge at the resonance conditions the neutron wave function density is resonantly enhanced in the channelling layer B [10] and hence all the neutron interactions such as the spin-flip cross section [11], the diffuse scattering [12,13], alpha-particles [14] or gamma-quanta [15] emission. 

\begin{figure}[ht]
       \includegraphics[clip=true,keepaspectratio=true,width=1\linewidth]{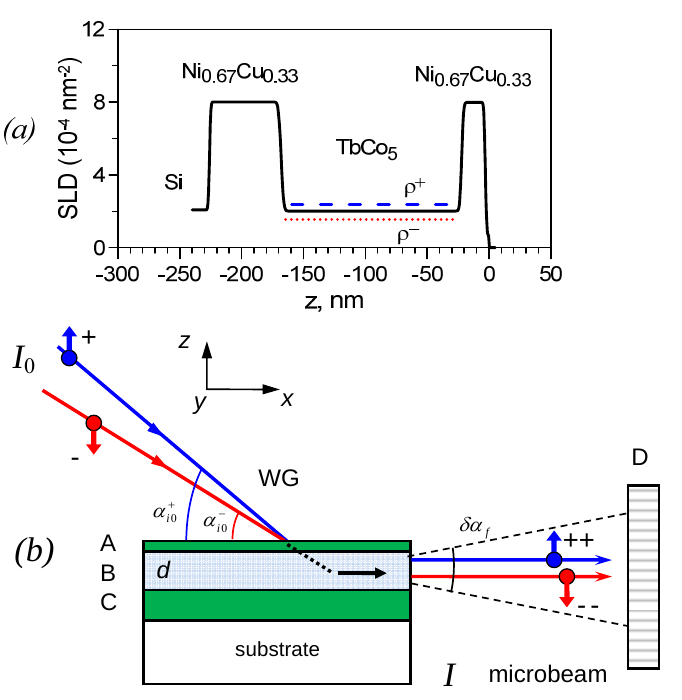}
          \caption{(a) The SLD depth profile of the investigated waveguide structure. (b) Geometry of the experiment: I0 is incident polarized neutron beam, WG is the waveguide, I is the guided microbeam outgoing from the sample edge, D is a position-sensitive detector.}
   \label{fig1}
 \end{figure}

Let $r_{BA (BC)}$  be the complex reflection amplitudes of the tunnelling layer and of the reflector layer, from inside out. Let $k_B=\sqrt{k_0^2-u \pm \mu B}$  be the normal component of neutron wave-vector in the investigated layer of thickness $d$. The complex reflection coefficients can be calculated numerically based on the parameters of the reflector layer, the tunnelling layer, and of the resonant layer. Within a weak magnetization approximation, their dependence on the magnetic induction can be neglected. The resonance condition is given by
\begin{equation}
   arg(r_{BA})+arg(r_{BC})+2 k_Bd=2\pi n                       
\label{eq1}
 \end{equation}

Where $r_{BA}$  and  $r_{BC}$ are the complex reflection amplitudes from the complete structures to the left and to the right of the resonant layer, respectively. From (1) it follows that if the channelling layer has a non-zero magnetization,  $k_B^+ > k_B^-$ and hence the resonances for spin-up and spin-down neutrons are split. The splitting value  $\Delta k_{res}^2=(k_{res}^+)^2-(k_{res}^-)^2$   is proportional to the induction $B$:
 \begin{equation}
\begin{split}
 \Delta k_{res}^2= & 2\mu B-\{ \frac{1}{4} [ arg(r_{BA}^+r_{BC}^+)^2 -arg(r_{BA}^-r_{BC}^-)^2] \\
& - [arg(r_{BA}^+r_{BC}^+) -arg(r_{BA}^-r_{BC}^-) ] \pi n\} /d^2 \\
= & 2 \mu B + O[ arg(r_{BA}r_{BC}]                       
\label{eq2}
\end{split}
 \end{equation}

Contrary to standard reflectivity where the measured signal is the specular reflectivity, in a channelling measurement, the measured intensity is the neutron flux exiting the edge of the sample out of the channel (along the horizon) (see Fig. 1b). The experimental measurement consists in measuring the neutron flux as a function of the incidence angle on the waveguide structure. The theory of neutron channeling was developed in [16]. Experimentally, the phenomenon of neutron channeling was observed in [8,17,18] and the channelling length was measured experimentally in [19,20]. It has also been demonstrated that it may be used for production of unpolarised [21] and polarised [22,23] neutron microbeams, which was used for the investigation of amorphous magnetic wires [24,25].

In this communication, we compare the magnetic characterization of a planar waveguide with a weakly ferrimagnetic guiding layer by two methods: polarized neutron reflectometry (PNR) and polarized neutron channeling (PNC).  We take as an example the compound TbCo5 which is a ferrimagnet that has a saturation magnetization of about 22 mT at room temperature [26]. This material is attractive for its use in 'hard - soft ferrimagnetic' spin-valve systems and for the development of ultrafast magnetic switching devices [27].

The investigated TbCo$_5$ layer with relatively small nuclear SLD is placed between NiCu layers having high SLD to create a waveguide structure (Fig. 1a). The sample structure is { Ta(3~nm)/Ni$_{0.67}$Cu$_{0.33}$(15nm)/TbCo$_5$(150nm) /Ni$_{0.67}$Cu$_{0.33}$(50nm) //Si(substrate)}. It was grown on thermally oxidized Si substrates by using the magnetron sputtering technique. The base pressure in the MAGSSY sputtering chamber was tuned to be lower than 5x10$^{-9}$ mbar. The partial Ar pressure during growth was set to a value of 1.5x10$^{-3}$ mbar. During preparation, the substrates were kept at room temperature in order to avoid any additional thermal inter-diffusion at the interfaces. The Tantalum layer on the top protects the structure against oxidation. 

The neutron experiments were carried out at the polarized neutron reflectometer NREX (reactor FRM II, TUM, Garching, Germany) at a fixed wavelength $\lambda = 4.26$~\AA  ($\delta\lambda/\lambda = 1.5$~\%) and with an horizontal sample plane. The angular divergence of the incoming beam was adjusted by two slits before sample (separated by a distance of 2000 mm). For the reflectometry (respectively the channelling experiments), the width of both slits were set to 1~mm (respectively 0.2 mm), thus providing an angular divergence of the incident beam of $\delta \alpha_i  = 0.5$~mrad (respectively 0.1 mrad). Two single magnetic supermirrors with a polarizing efficiency of 99\% in transmission mode were used as the polarizer and the analyzer. The scattered neutrons were measured with a 200x200~mm$^2$ two-dimensional $^3$He position-sensitive detector with a spatial resolution of 3 mm placed at a distance 2400 mm from the sample position. The experiment was performed at T = 300 K in a magnetic field $\mu_0 H = 0.45$~T applied parallel to the surface and normal to the scattering plane. 

In Fig. 1b, the geometry of the  experiment is shown schematically. The incident neutron beam of intensity $I_0$ enters the surface of the waveguide (WG) under a glancing angle $\alpha_i$ . Neutrons tunnel through the thin upper layer into the guiding layer (or channel) of width $d$, are reflected from the bottom thick layer and channel in the guiding layer in the direction parallel to the sample surface. The neutron wave density is resonantly enhanced for some specific incident glancing angles $\alpha_{i,n}$ , where $n=0, 1, 2, ...$ is the resonance order. The neutrons channel inside the guiding layer over a distance of several millimetres (which corresponds to the channeling length $\xi_e$) and leave the channel end as a microbeam of intensity $I$. The angular divergence of the microbeam  $\delta \alpha_f$ is defined by the Fraunhofer condition of diffraction on a narrow slit which is the exit of the channel of width $d$. For a neutron wavelength $\lambda$, the angular divergence of the microbeam is $\delta \alpha_f \sim \lambda /d$ . The microbeam intensity is recorded by a position-sensitive detector (D) around the direction of the sample plane (sample horizon).

The PNR curves (Fig. 2a) were measured under a magnetic field of 0.45 T in the range of momentum transfers $Q=0.06\div1$~(nm$^{-1}$). The two non-spin-flip $R^{++}$  and  $R^{--}$ reflectivities were measured within 12 hours. The non spin-flip signals show a weak spin asymmetry $A=(R^{++}-R^{--})/(R^{++}+R^{--})$  which is smaller than 0.1 (Fig. 2b). No measurable signal was detected in the spin-flip channels which confirms that the sample was fully saturated. The fit of the PNR curves led to the SLD profile shown in Fig. 1a). The measured spin asymmetry of the NSF channels can be reproduced with an induction $B = 20$~mT, but the fit procedure may lead also to significantly different results. 

\begin{figure}[ht]
       \includegraphics[clip=true,keepaspectratio=true,width=1\linewidth]{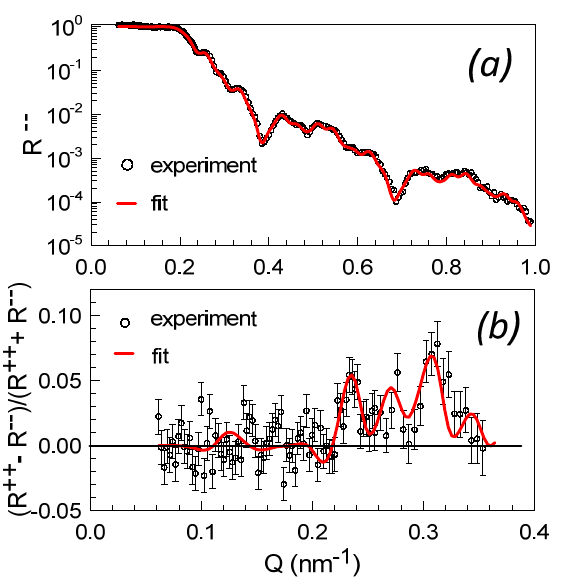}
          \caption{(a) Experimental (dots) and model (solid line) reflectivity for \'--\' channel. (b) Experimental (dots) and model (solid line) neutron spin asymmetry.}
   \label{fig2}
 \end{figure}

\begin{figure}[ht]
       \includegraphics[clip=true,keepaspectratio=true,width=1\linewidth]{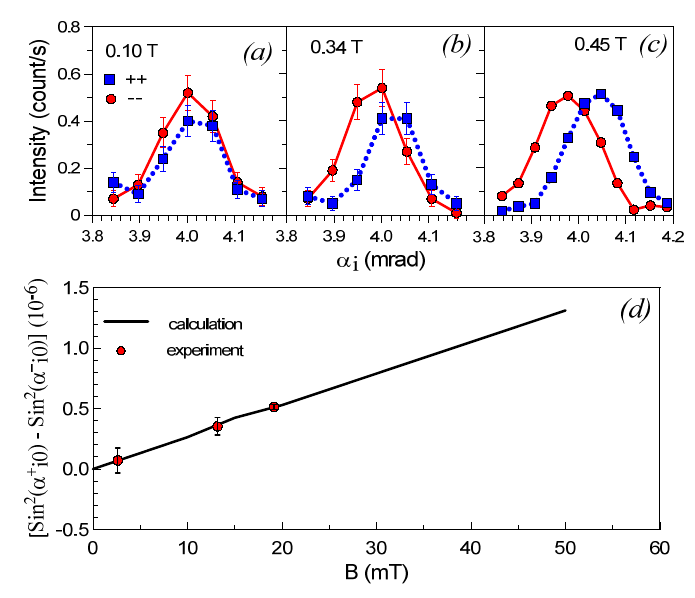}
          \caption{(a) Neutron microbeam intensity for spin states \'++\' (squares) and \'--\' (circles) as a function of the incidence glancing angle at the applied magnetic field: (a) 0.10; (b) 0.34 and (c) 0.45 T. (d) Splitting of the resonance glancing angles as a function of the magnetic induction (the line is calculation and the symbols are experiment).}
   \label{fig3}
 \end{figure}

In a second experiment, the neutron channelling signal was measured. The neutron microbeam intensity of the resonance $n=0$ extracting background level is presented in Fig. 3a-c
as a function of the incidence glancing angle (closed symbol for '++' and open symbols for '$--$'). The {\it signal/background} ratio is about 3.

The error bars for the field 0.45 T (Fig. 3c) are smaller than the symbol size. One can see that the angular splitting of the resonance peaks for '++' and '$--$' microbeams is well defined for the fields 0.35 T and 0.45 T.
For example, for the external filed 0.45 T (Fig. 3c)   providing a magnetically saturated state for the layer, it is possible to precisely fit the positions of the resonance incidence angle $\alpha^+_{i0} = 4.04$~mrad and $\alpha^-_{i0} = 3.98$~mrad with accuracy better than 0.001 mrad. Based on these angles, it is possible to estimate the value of the magnetic induction $B$ by writing the equation (2) as function of the resonant grazing angles: 

 \begin{equation}
\begin{split}
 B=\frac{\hbar^2}{4 \mu m}(\frac{2\pi}{\lambda})^2  [\sin^2(\alpha^+_{i0})-\sin^2(\alpha^-_{i0}]- O[ arg(r_{BA}r_{Bc}]         
\label{eq3}
\end{split}
 \end{equation}

where $\hbar$ is the Plank's constant, $m$ is the neutron mass, $\lambda$ is the wavelength of the neutrons, and $\mu$ is the neutron magnetic moment.

The result is $B=19.53$~mT when neglecting the argument factors $O[ arg(r_{BA}r_{BC}]$. Using the calculated arguments, the result becomes: $B=19.56$~mT. As a consequence, by neglecting completely the argument factors, the value of the magnetic induction is determined within a 10$^{-3}$ precision.

The calculated difference of squared sinuses of the resonance angles as a function of $B$ is presented in Fig. 3d as a solid line. The symbols corresponds to the experimental values of the microbeam peaks positions. The error bars were defined by the peaks positions fitting. The extracted experimental value of $B$ is in agreement with previous measurements by SQUID magnetometer[26]. In addition to the saturation state we have also measured the resonance splitting for intermediate magnetic states which occur during the magnetization reversal for external fields of  0.1 T and 0.34 T. These data does further demonstrate the direct sensitivity of the method to the averaged weak magnetization values. 

In conclusion, we propose to use another method for measuring of the absolute values of the magnetization of thin films. The method is based on the measurement of the difference of resonance positions for channeled spin-up and spin-down neutrons through a waveguide system with the investigated magnetic film in the middle. Compared to conventional PNR [28], the method does not require any modeling of the data and hence is more robust. It is also very accurate (0.1 mT resolution).

In the specular reflection below the critical angles, the spin asymmetry does not exhibit a sizable finite value because of flux conservation. On the other hand, in the case of the neutron channeling effect, there are two factors of neutron interaction enhancement:
\begin{itemize}
\item {At the resonances, the time of neutron interaction with matter is increased due to multiple reflections inside the resonator in the direction perpendicular to the sample surface.}
\item {The critical condition for the resonances is different for the '+' and '-' spin states which is very sensitive to the magnetization absolute value.}
\end{itemize}

The studied system was in a state where the magnetization was collinear with the applied field for the highest applied external field. In the case of non-collinear magnetization which most likely occurs for the intermediate magnetic states, one would have to apply polarization analysis to resolve the spin orientations, taking carefully into account polarization imperfections of the polarizer and analyzer. This may lead to rather critically small signal/background ratios which may require higher neutron flux for obtaining quantitative results. Nevertheless, spin-flipped neutrons in resonators were observed in model experiments [6,8,9] and were used for the investigations of proximity effect in ferromagnetic/superconductor system [29]. One may also reduce the background by using the beam-splitting effect in external fields for non-collinear arrangements. For example, in [5,6,30] beam-splitting was used to observe standing waves in off-specular region with higher ratio {\it signal/background}. In [24,25] a ratio {\it signal/background}$\approx$10 for the microbeam outgoing from a planar waveguide. Thus, the method of polarized neutron channeling can be used for the investigations of weak magnetism (collinear or non-collinear) in thin films.

Concluding, a tri-layer waveguide structure with weakly magnetic guiding layer TbCo$_5$ was studied using polarized neutron channeling. Due to the resonant nature of the neutron channeling and the magnetic splitting of the resonances, the polarized neutron microbeam method is more sensitive than the polarized neutron reflectometry. Thus, the neutron channeling can be used as a new method for the investigation of weakly magnetic layers with magnetization on the order of 10 mT.

This work is based upon experiments performed at the NREX instrument operated by Max-Planck Society at the Heinz Maier-Leibnitz Zentrum (MLZ), Garching, Germany. The neutron part of the project has been supported by the European Commission under the 7th Framework Programme through the "Research Infrastructures" action of the Capacities Programme, NMI3-II, Grant Agreement number 283883. Also this work has been financially supported by the French project IMAMINE 2010-09 T.


\end{document}